\documentclass[preprint2]{aastex}

\usepackage{fullpage}
\usepackage{bm}		%

\usepackage{booktabs}
\usepackage{latexsym}
\usepackage{amsmath}
\usepackage{amssymb}
\usepackage{ascmac}
\usepackage{graphicx}
\usepackage{theorem}
\usepackage{color}
\usepackage{subfigure}
\usepackage{float}

\newcommand{\bfm}[1]{\mbox{\boldmath{$#1$}}}

\shortauthors{Hirabayashi et al.}

\begin{document}


\title{Analysis of Asteroid (216) Kleopatra \\ using dynamical and structural constraints}


\author{Masatoshi Hirabayashi}
\affil{Aerospace Engineering Sciences, 429 UCB, University of Colorado, Boulder, CO 80309-5004 United States}
\email{masatoshi.hirabayashi@colorado.edu}

\and
\author{Daniel J. Scheeres}
\affil{Aerospace Engineering Sciences, 429 UCB, University of Colorado, Boulder, CO 80309-0429 United States}





\begin{abstract}
Given the spin state by \cite{Magnusson1990}, the shape model by \cite{Ostro2000}, and the mass by \cite{Descamps2011}, this paper evaluates a dynamically and structurally stable size of Asteroid (216) Kleopatra. In particular, we investigate two different failure modes: material shedding from the surface and structural failure of the internal body. We construct zero-velocity curves in the vicinity of this asteroid to determine surface shedding, while we utilize a limit analysis to calculate the lower and upper bounds of structural failure under the zero-cohesion assumption. Surface shedding does not occur at the current spin period (5.385 hr) and cannot directly initiate the formation of the satellites. On the other hand, this body may be close to structural failure; in particular, the neck may be situated near a plastic state. In addition, the neck's sensitivity to structural failure changes as the body size varies. We conclude that plastic deformation has probably occurred around the neck part in the past.  If the true size of this body is established through additional measurements, this method will provide strong constraints on the current friction angle for the body. 
\end{abstract}





\section{Introduction}

Asteroid (216) Kleopatra, classified as a M-type in the \cite{Tholen1984} taxonomy or as a Xe-type in the \cite{Bus2002} taxonomy, has been of interest for a few decades due to its odd shape and fast spin rate. It orbits in the main belt, and asteroids of this type have not been targeted yet; therefore, this asteroid is not well understood. 

\subsection{Observational studies for Asteroid (216) Kleopatra}
However, we have significant information on (216) Kleopatra including the shape (\citealt{Ostro2000}), the spin period (5.385 hr, \citealt{Magnusson1990}), and the mass (4.64$\times$10$^{18}$ kg, \citealt{Descamps2011}). Not as certain, and subject to different interpretations, is its total size. 

Past researches have investigated this asteroid's shape by different observation techniques. From lightcurve observations, \cite{Scaltriti1978} confirmed shape elongation of this asteroid and small differences of the magnitudes at the maxima and the minima (see Fig.4 in their paper). They pointed out that those differences came from either different reflectivity or a shadowing effect. \cite{Weidenschilling1980} estimated (624) Hektor as a binary asteroid of which components are nearly in contact, considering a hydrostatic stable equilibrium shape. Then, he applied this technique to (216) Kleopatra, which has a (624) Hektor-like lightcurve, and showed that an amplitude of 3.3 slightly exceeds the value of a contact binary, but a contact binary model with its spin period recovers a reasonable density of 3.9 g/cm$^3$. Lightcurve observations by \cite{Zappala1983} revealed that a triaxial ellipsoid model fits their observations. On the other hand, \cite{Cellino1985} found that a binary model is compatible with their lightcurve data. They also pointed out that the amplitude 0.9 mag by \cite{Zappala1983} is an estimation, while the amplitude of this asteroid highly depends on the phase. Occultations by \cite{Dunham1991} estimated the size as dimensions of 230 km by 55 km. Furthermore, \cite{Mitchell1995} performed radar observations; however, although they obtained the Kleopatra echoes which are similar to those of bifurcated asteroid (4769) Castalia, their coarse data set precluded them from determining the shape. They also attempted to detect the shape from occultation data; however, since the model used was simplistic, they could obtain no evidence for a bifurcation. 

From comprehensive radar observations, \cite{Ostro2000} constructed a three-dimensional bi-lobed polyhedral shape model with dimensions of 217 km by 94 km by 81 km, although they indicated that the absolute size uncertainty was up to 25$\%$. On the other hand, from the Fine Guidance Sensors (FGS) aboard HST, \cite{Tanga2001} confirmed that their shape model is consistent with radar observations by \cite{Ostro2000}. 

Later studies compared their observation analyses with the \cite{Ostro2000} model. \cite{Hestroffer2002A} showed that a larger and more elongated model is consistent with the occultations, the photometric and the interferometric HST/FGS results. Adaptive optics observations by \cite{Hestroffer2002B} showed that their model is consistent with the \cite{Ostro2000} shape model, although these observations could not rule out the possibility that this asteroid is a binary asteroid. \cite{Takahashi2004} performed lightcurve simulations based on the binary model by \cite{Cellino1985}, the contact binary model by \cite{Tanga2001}, and the polyhedron model by \cite{Ostro2000} to report that while the binary model and the contact binary model fit their lightcurve simulations, the \cite{Ostro2000} shape model could not. It is worth noting that for simulations using the \cite{Ostro2000} shape model, they used the size estimated by \cite{Ostro2000}, which may be smaller than the actual size. 

From near-infrared adaptive optics observations, \cite{Descamps2011} calculated the mass as 4.64$\times$10$^{18}$ kg from the mutual gravity interaction between (216) Kleopatra and its satellites. \cite{Kaasalainen2012} attempted to construct a new shape model, using multiple observation data (photometry, adaptive optics, occultation timings, and interferometry); however, they mentioned that the data were not compatible and thus further analyses are necessary. We look forward to their new shape model. 

\begin{figure}[p!]
\begin{center}
\includegraphics[width=3in]{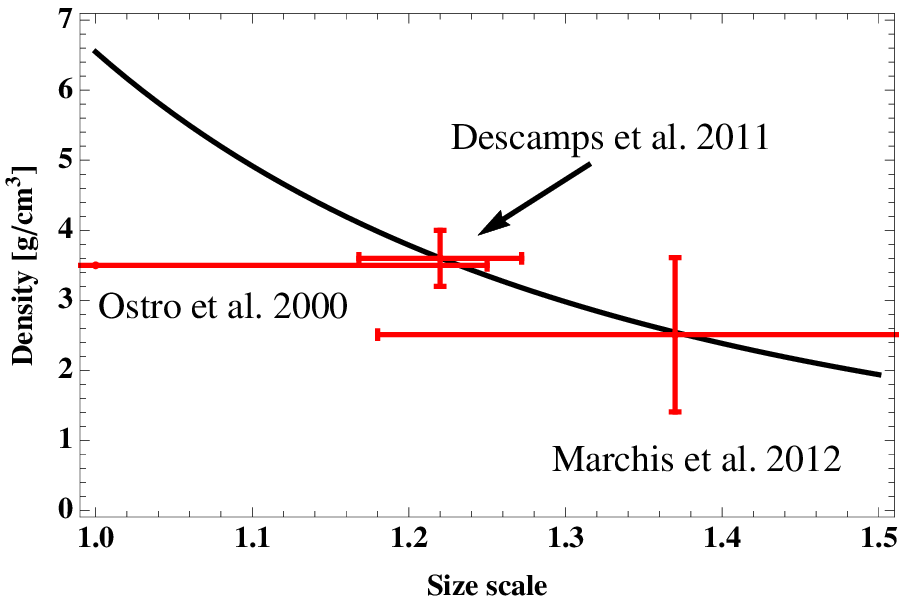}
\caption{Relation between the size scale and the density. Given the mass $4.64\times10^{18}$ kg by \cite{Descamps2011} and the shape by \cite{Ostro2000}, the curve describes the ideal density as a function of the size scale. The actual density and the size scale should be on this curve. The error bars are observation values by \cite{Ostro2000}, \cite{Descamps2011}, and \cite{Marchis2012}. Note that Marchis et al. (2012) did not take into account the shape modification for their estimation (personal communication, Marchis, 2013). }
\label{Fig:densityScale}
\end{center}
\end{figure}

\cite{Ostro2000} estimated the equivalent diameter\footnote{An equivalent diameter is a diameter of a sphere with the same volume as the shape.} as 108.6 km by radar observations and the surface bulk density as 3.5 g$/$cm$^3$ from the surface reflectivity. Note that the latest version of the shape model provides a mass of $7.09 \times 10^{5}$ km$^3$, which is equal to an equivalent diameter of $111.1$ km. \cite{Tedesco2002} reported the IRAS equivalent diameter as 135.07 km, while the estimation by \cite{Descamps2011} is consistent with the \cite{Tedesco2002} size. On the other hand, from observations with Spitzer/IRS, \cite{Marchis2012} derived the equivalent diameter as 152.5 km by using the Near-Earth Asteroid Thermal Model (we referred to Table 5 in their paper). These researches imply that the size of (216) Kleopatra has not been well understood. Figure \ref{Fig:densityScale} shows the comparison between the estimated size scale and the bulk density. Scale size means an equivalent diameter relative to that of the \cite{Ostro2000} size, i.e., $7.09 \times 10^5$ km$^3$. The \cite{Ostro2000} size is 1.00, the \cite{Descamps2011} size is 1.22, and the \cite{Marchis2012} size is 1.37. For \cite{Ostro2000}'s estimation, we only show the error bar of the size scale (the region of the horizontal axis). Note that the \cite{Ostro2000} density is based on their surface reflectivity estimation.

From these papers, it seems that the \cite{Ostro2000} shape has been confirmed by other researches, but that the size has not. This study, therefore, utilizes the \cite{Ostro2000} shape and keeps its size as a free parameter. Note that although our model provides insight on a surprisingly important role that the total size is related to its stability, future measurements will be able to provide real insight on the strength of this body by using our current analysis. 

\subsection{Theoretical studies of internal structure of a rotating ellipsoid}
Theoretical studies of internal structure of a rotating triaxial ellipsoid have been of interest for a long time. In particular, the elastic stresses have been discussed for more than 100 years. \cite{Chree1889} provided a complete elastic solution for a rotating sphere in terms of polar coordinates, although \cite{Chree1891} pointed out that an application of the \cite{Chree1889} theory to the Earth may be limited because the dependency of the initial internal stress state on the history causes difficulty in determining the ellipticity. \cite{Love1944} discussed this point in detail\footnote{See Article 75 and 176 in \cite{Love1944}.} and avoided this difficulty by assuming that a rotating body is homogeneous and incompressible. \cite{Dobrovolskis1982} provided an elastic solution in Cartesian coordinates with nearly incompressible Poisson's ratio $\nu=0.46$, but gave the comment that this Poisson's ratio eases the difficulty for the sudden collapse of the surface due to turning on the self-gravity, but this situation may be unrealistic. \cite{Washabaugh2002} focused on elastic energy of a rotating ellipsoid. They found that the stress state in the compressible case can relax around the lowest elastic energy point more easily than that in the incompressible case. \cite{Kadish2008}, on the other hand, investigated internal elastic-stresses of a uniformly rotating self-gravitating accreted ellipsoid. 

On the contrary, approaches using plastic theory are relatively new. Deriving a general solution of the internal stress on the zero-cohesion assumption with regard to the Mohr-Coulomb (MC) yield criterion, \cite{Holsapple2001} found that for an uniformly rotating ellipsoid, the upper limit load and the lower limit load are identical. \cite{Holsapple2004} proposed new definitions of local stability of shapes and applied this stability condition to a uniformly rotating solid ellipsoid. On the other hand, in the case of a rod, a disk, and an ellipsoid, \cite{Holsapple2008A} confirmed that an actual failure occurs between the lower and upper limit loads. \cite{Holsapple2010} derived deformation paths of an ellipsoid due to a YORP-induced spin. His result was consistent with numerical simulations by \cite{Sanchez2012}. Using plastic material condition with regard to the Drucker-Prager yield criterion, \cite{Sharma2010} formulated equilibrium shapes of rubble-pile binaries and investigated the current material properties of contact binary asteroids. In this experiment, he modeled (216) Kleopatra as a contact binary and mentioned, ``We will model Kleopatra as a congruent contact binary with prolate ellipsoidal members. We thus ignore the `bridge' connecting the binary's members, thereby assuming that the bridgeÕs internal strength and mass are negligible."

\subsection{Outline of the present study}
Our goal here is to investigate the failure mode that (216) Kleopatra most likely experiences and to evaluate a structurally stable size, in the hope of finding constraints on and predictions of what its current size should be. The spin rate by \cite{Magnusson1990}, the shape model by \cite{Ostro2000}, and the mass by \cite{Descamps2011} are considered to be constant properties, while the size is varied for determining the stable size. This paper discusses two possible failure modes: material shedding from the surface and structural failure due to plastic deformation. 

The surface shedding condition is given by using zero-velocity curves. This technique allows us to visualize allowable and non-allowable regions of orbital motion of a massless particle about the primary and its dynamical equilibrium points. It is also interesting that surface shedding initiates the formation of the satellites. According to \cite{Descamps2011}, the satellites are formed by material shedding. We develop a model used for predicting whether or not these satellites result from surface shedding. 

On the other hand, the structural failure condition is discussed by a limit analysis. Note that a limit analysis explicitly assumes that materials are elastic-perfectly plastic, the yield envelope is smoothly convex, the material behavior follows an associated flow rule, and the velocity fields are homogeneous. This technique gives the lower and upper bounds of structural failure. The lower bound is the condition where a body does not experience plastic collapse. We obtain this bound by solving elastic solutions with commercial finite element software ANSYS (Academic Teaching Introductory, 14.0). On the other hand, the upper bound is the condition where a body must fail plastically due to its limit load. We calculate this bound for a whole volume and that for a partial volume. To determine plastic failure of a partial volume, we assume that (216) Kleopatra is symmetric about the principal axes.

This paper is organized as follows. The first part summarizes physical properties used in this analysis. The second part defines surface shedding and structural failure. Third, this paper introduces the techniques for determining these failure modes. Then, the surface shedding condition and the failure condition are compared. In particular, the satellites formation and the stable size estimation are the prime results in this paper. 

\section{Known physical parameters of Asteroid (216) Kleopatra}
\begin{figure}[p!]
\begin{center}
\includegraphics[width=3in]{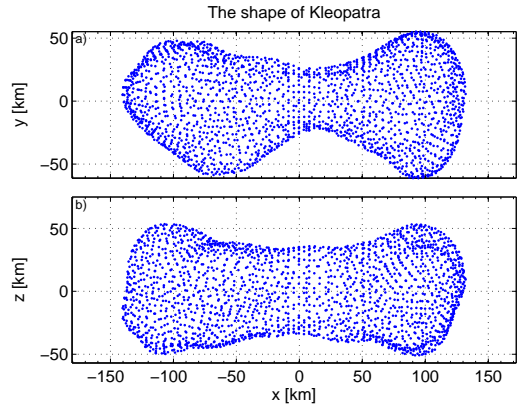}
\caption{The projection of (216) Kleopatra's shape model by \cite{Ostro2000}. This shape model consists of surface points (vertices) and the order of surface elements (faces). The plot shows the vertices in Cartesian coordinate frame. The upper plot shows the projection onto the $x-y$ plane, while the lower plot is that onto the $x-z$ plane. Note that $x$, $y$, and $z$ lie along the minor, intermediate, and major principal axis, respectively. The size scale of these plots is adjusted so as to be the same as 1.22, i.e., the \cite{Descamps2011} size.}
\label{Fig:Kleopatra_Shape}
\end{center}
\end{figure}

The spin period is fixed as 5.385 hr (\citealt{Magnusson1990}). On the other hand, we utilize the shape model by \cite{Ostro2000}. Figure \ref{Fig:Kleopatra_Shape} shows the projection of the surface points onto the $x-y$ and $x-z$ plane, where $x$, $y$, and $z$ are the minor principal axis, the intermediate principal axis, and the major principal axis, respectively. The size in this figure is the same as the estimation by \cite{Descamps2011}. The bifurcation structure can be confirmed in the middle of the body. Furthermore, the mass is fixed as the \cite{Descamps2011} mass, i.e., $4.64\times10^{18}$ kg. Those properties are described in Table \ref{Table:Kleopatra}. This analysis assumes density homogeneity and uniform rotation. In addition, physical parameters of the satellites are given by \cite{Descamps2011} (see Table \ref{Table:satellite}). To discuss dynamics of the satellites, we assume that the orbital planes of the satellites are parallel to the equatorial plane of (216) Kleopatra. This result comes from their hypothesis; however, it finally allowed them to derive a relevant orbit solution. Moreover, from the observation of a stellar occultation by (216) Kleopatra in 1980 (see Sec.3.4 in \citealt{Descamps2011}),  they could interpret the reported secondary event from their simple solution (Descamps, 2013, personal communication).

Material properties are also critical parameters to characterize the body behavior, although they are usually unknown. It is assumed that materials considered here are elastic-perfectly plastic, the yield envelope is smoothly convex, the material behavior follows an associated flow rule, and the velocity fields are homogeneous. Those assumptions are crucial in using a limit analysis. Here, the Mohr-Coulomb (MC) yield criterion, a function of a friction angle and cohesion, is utilized to describe (216) Kleopatra's yield condition. This study assumes zero-cohesion. \cite{Holsapple2007} reported that large bodies such as (216) Kleopatra are located in the gravity regime in which cohesion is negligible. 

\section{Possible failure modes}
\subsection{Surface shedding}
\begin{figure}[p!]
\begin{center}
\includegraphics[width=3in]{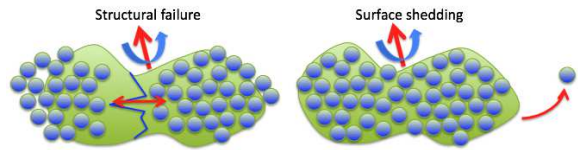}
\caption{Structural failure (left) and surface shedding (right).}
\label{Fig:Reshaping}
\end{center}
\end{figure}

Surface shedding is a dynamical-oriented process of small particles on the surface. The process occurs due to the balance of the total forces acting on these particles. When the spin rate is above this condition, any particles that experience outward forces fly off and do not immediately come back to the surface (see the right side in Fig.\ref{Fig:Reshaping}). At this point, the internal body should be below structural failure. Note that the necessary condition for loose material to fly off the surface is that {\it dynamical equilibrium points reach the surface}. \cite{Guibout2003} gave the similar analysis for the surface stability of a uniformly spinning  ellipsoid. Their results showed that as the spin period increases gradually, the stable regions move to the extremities along the minor principal axis. Finally, the saddle equilibrium points touch the surface and the stable regions disappear from the extremities. Throughout the text, we call this condition the ``first shedding". 

Another curious question is the origin of the satellites. \cite{Descamps2011} estimated the physical properties of the satellites (Table \ref{Table:satellite}). They also stated, ``as to the origin of Kleopatra's companions, they could be a by-product of the spinning-up process leading to mass shedding in orbit." Is it possible that surface the materials were shed to form the satellites in the past? To answer this question, we perform a simple analysis, using the necessary condition of the first shedding. 


\subsection{Structural failure}
We define term ``structural failure" as the end condition where plastic flow spreads over a target volume given by arbitrary cuts. The volume experiencing structural failure has subsequent deformation after the first yield. This statement implies that plastic flow appearing in a small volume does not always lead structural failure. In such a case, when this element is unloaded, it is sustained by residual stress (\citealt{Chakrabarty2006}). The present study focuses on structural failure of a slice perpendicular to the minor principal axis, because the centrifugal force causes the strongest tension and shear over this volume. At the structural failure condition, the slice initiates a catastrophic break-up of a body, which is depicted on the left side in Fig.\ref{Fig:Reshaping}. Note that the situation where a body is separated into two components may need further angular momentum gain from structural failure.


\subsection{Relation between surface shedding and structural failure}
Past researches reported that surface shedding and structural failure are highly correlated in terms of shapes and material properties. \cite{Eriguchi1982} pointed out that a symmetrically incompressible fluid becomes a dumbbell shape before a break-up, while an asymmetrically incompressible fluid becomes pear-shaped before mass shedding. Using the $N$-body code, \texttt{pkdgrav}, \cite{Walsh2008} and \cite{Walsh2012} demonstrated that mass shedding from a spherical aggregate initiates a binary system. Their analysis also revealed that the (66391) 1999 KW4 type equatorial ridge may be formed by landslides. On the other hand, using a Soft-Sphere Discrete Element Code, \cite{Sanchez2012} reported that particle-particle surface friction makes a spherical shape experience surface shedding near its equatorial plane and makes an ellipsoidal shape break into two components.


\section{Analysis Method}

\subsection{Determination of the first shedding}
We note that a classic technique of this type is to compute the effective gravity slope, which is the direction of the total force (usually the gravitational acceleration and the centrifugal acceleration) from the local downward normal. For example, \cite{Ostro2006} visualized the effective gravity slope of (66391) 1999 KW4. This technique is useful when one wants to focus on the surface condition of general shapes. However, the capability of this technique is limited if one wants to track the orbital motion in the vicinity of the primary. 

This paper applies zero-velocity curves to the determination of the first shedding. Again, the necessary condition for the first shedding is equivalent to that one of the equilibrium points reaches the surface first. The zero-velocity curves allow us to visualize both the equilibrium points and the constraints on the motion. However, for the use of this method, there are some cautions that should be noted. 

First, this computation only considers the balance between the gravitational acceleration and the centrifugal acceleration. In other words, some attractive and repulsive forces that may be significant in an asteroid's environment are not involved in this analysis. For example, a cohesive force is one of effective attractions that may be significant for a small asteroids. On the contrary, (216) Kleopatra is assumed to be zero-cohesion, so the zero-velocity curve computation gives reasonable estimations for the first shedding. Second, even when the spin state is below the first shedding, there is a phenomenon that materials may be ejected in space by some reasons, e.g., saltation by landslides. The present technique does not consider this phenomenon. As given earlier, surface shedding is the process that any particles experiencing zero-forces are about to fly off and do not immediately come back to the surface after lifting off. This condition is different from the condition where particles are simply ejected from the surface. Particles ejected below the first shedding usually come back to the surface immediately and do not contribute to major collapse processes. Third, we assume that the original shape does not change. This assumption implicitly makes the first condition more conservative than the actual first shedding because shape deformation allows the equilibrium points to touch the surface at a slower spin period. Consideration of this effect is beyond the present paper. 

\subsubsection{Zero-velocity curves}
The motion of a massless particle affected by the gravity from the primary in the rotating frame is described as
\begin{eqnarray}
\ddot x - 2 \omega \dot y &=& - U_x + \omega^2 x, \nonumber \\
\ddot y + 2 \omega \dot x &=& - U_y + \omega^2 y , \label{Eq:dynamics}\\
\ddot z &=& - U_z, \nonumber 
\end{eqnarray}
where $U$ is the potential and the subscripts of $U$ mean the partial derivative with respect to the position. The $x$ axis, the $y$ axis, and the $z$ axis lie along the minor principal axis, the intermediate principal axis, and the major principal axis, respectively.

The Jacobi integral $C_J$ is given multiplying each equation of Eq.(\ref{Eq:dynamics}) by $x$, $y$, and $z$, summing these equations, and integrating: 
\begin{eqnarray}
C_J=\omega^2 (x^2+y^2)-2 U - \dot x^2 - \dot y^2 - \dot z^2. \label{Eq:JacobiIntegral}
\end{eqnarray}
If $\dot x = \dot y = \dot z =0$, then Eq.(\ref{Eq:JacobiIntegral}) becomes
\begin{eqnarray}
C_J\le\omega^2 (x^2+y^2)-2 U.  \label{Eq:ZV}
\end{eqnarray}
Equation (\ref{Eq:ZV}) gives constraints on the motion of the particle. The closed boundary defined by $C_J=\omega^2 (x^2+y^2) - 2 U$ is the zero-velocity curve. It also describes the force that points normal to them and towards the allowable regions. In general cases, there exist more than four equilibrium points especially when the primary is nearly spherical. However, because of its highly bifurcated shape, (216) Kleopatra only has four equilibrium points: two of them sitting along the minor principal axis (saddle points) and the other two lying around the intermediate principal axis (center points). When one of the equilibrium points first touches the surface\footnote{The saddle point always reaches the surface first.}, the condition is called the first shedding condition. 

\cite{Yu2012, Yu2013} investigated the zero velocity curves and the equilibrium points of (216) Kleopatra, using the same technique that we showed above. We emphasize that our computation results are different from their results by the following reason. This paper uses a constant mass of $4.64\times10^{18}$ kg based on comprehensive observations by \cite{Descamps2011}. On the other hand, although \cite{Yu2012, Yu2013} stated that \cite{Descamps2011} obtained an accurate mass, they utilized the estimations by \cite{Ostro2000}, i.e., a volume of $7.09\times10^5$ km$^3$ and a density of 3.6 g/cm$^3$. In Appendix \ref{App:Comp}, we describe computational comparison between \cite{Yu2012,Yu2013} and our calculation.

\subsubsection{Numerical search for the first shedding}
The numerical algorithm by \cite{Werner1996} will be used to calculate accurate external gravity forces of a polyhedoral model. The following is the numerical scheme. First, given a spin period, we compute the zero-velocity curves. Second, we explicitly calculate the equilibrium points in each iteration. If the zero-velocity curve at the energy level of the saddle points touches the surface, the iterative scheme stops; otherwise, the spin period is updated to be faster. At the same energy level, the saddle points usually reach the surface first. Since the size is a free parameter, our code searches for this spin period in the test size scale range (we will show the range later). 

\subsection{Determination of the upper and lower bounds of structural failure}

\subsubsection{The Mohr-Coloumb yield criterion of cohesionless materials}
\begin{figure}[p!]
\begin{center}
\includegraphics[width=3in]{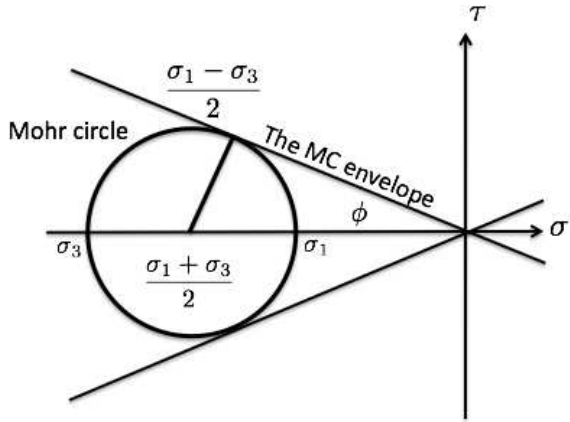}
\caption{Mohr-Coloumb yield envelope identical to the slope touching a Mohr circle. The inclination depends on a material's properties. If the stress state is within the envelope, only elastic deformation occurs. On the other hand, the stress is on the envelope, a body experiences plastic strain.}
\label{Fig:MC}
\end{center}
\end{figure}

The Mohr-Coulomb (MC) yield criterion for cohesionless materials is given as
\begin{eqnarray}
g(\sigma_1, \sigma_3, \phi) \le 0, \label{Eq:MC}
\end{eqnarray}
where 
\begin{eqnarray}
g(\sigma_1, \sigma_3, \phi) = \frac{\sigma_1 - \sigma_3}{2} \sec \phi + \frac{\sigma_1 + \sigma_3}{2} \tan \phi.  \label{Eq:g}
\end{eqnarray}
$\phi$ is the angle of internal friction. The principal stresses are denoted by $\sigma_i \: (i=1,2,3)$, where $\sigma_3 < \sigma_2 < \sigma_1$. The MC envelope is identical to the slopes touching the Mohr circle centered at $(\sigma_1+\sigma_3)/2$ with a radius of $(\sigma_1-\sigma_3)/2$ (see Fig.\ref{Fig:MC}) in the $\sigma$-$\tau$ space, where $\sigma$ is the normal stress and $\tau$ is the shear stress. The slopes go through the origin, if materials are zero-cohesion. The angle between the slope and the $\sigma$ axis is identical to $\phi$. The elastic region is inside the yield envelope, while the plastic region is on the envelope\footnote{This comes from our assumption of elastic-perfectly plastic materials.}.  

In the three-dimensional principal stress space, the MC yield envelope is a hexagonal cone opening to the negative direction along the hydrostatic pressure and is not smooth at the tension and compression meridians. The smoothness of the yield envelope plays a role in a limit analysis technique. However, since our interest is to investigate real shapes, we barely encounter the stress state at these meridians.

\subsubsection{Computation of body forces}

The gravitational acceleration and the centrifugal acceleration are calculated by decomposing the original body into smaller elements: cubes (inside of the body) and polygons (on the surface). On the assumption that the density is constant and each element is so small that we can use a simple inverse-square law for spheres, the gravitational acceleration of element $s$ is described as 
\begin{eqnarray}
{\bfm b}_{sg} = - G \rho \sum_{t \ne s} \frac{V_t}{r^3_{st}} ({\bfm r}_s - {\bfm r}_t), \label{Eq:GravForce}
\end{eqnarray}
where $\rho$ is the density, $G$ is the gravitational constant, $V$ is the volume of an element, and element $t$ does not overlap $s$. ${\bfm r}$ is a position vector from the origin to an element and $r$ is the Euclidean norm of ${\bfm r}$. The centrifugal acceleration is described as
\begin{eqnarray}
 {\bfm b}_{sc} = - {\bfm \Omega} \times {\bfm \Omega} \times {\bfm r}_s = 
 \Omega^2
 \begin{bmatrix}
 x_{s} \\ y_{s} \\ 0
 \end{bmatrix}.
\end{eqnarray}
where $\bfm \Omega$ is the spin vector $\Omega [0, 0, 1]^T $. The total body force vector is now given as
\begin{eqnarray}
{\bfm b}_{s} =  {\bfm b}_{sg} +  {\bfm b}_{sc}.
\end{eqnarray}
The following discussion will use scalar notations $b_i$, $(i=1,2,3)$, for the component of the body force vector, instead of vector notation ${\bfm b}_s$. For computation of the stresses, $b_i$ is substituted into the equilibrium equation, which is given as 
\begin{eqnarray}
\frac{\partial T_{ij}}{\partial x_j} + \rho b_i = 0,
\end{eqnarray}
where $T_{ij}$ is a stress component in Cartesian coordinates 

\subsubsection{Limit analysis}
Limit analysis is a technique for calculating plastic collapse load at which an idealized body deforms without limit. At plastic collapse loading, a body should deform and fail plastically. Here, an idealized body means that a body is characterized by (i) elastic-perfectly plastic materials, (ii) convex yield criterion, (iii) an associated flow rule, and (iv) homogenous velocity fields. The last idealization allows us to apply average-stress techniques (shown below) to determination of the upper bounds by a limit analysis. The further details of limit analysis can be found in \cite{Chen1988} and \cite{Chakrabarty2006}.  

Let us discuss the definition and computation of the lower bound. According to \cite{Chen1988}, the lower bound theorem states, ``if an equilibrium distribution of elastic stress can be found which balances the body force in a specific volume and the applied loads on the stress boundary and is everywhere below yield, then the body at the loads will not collapse." This theorem is interpreted as the condition where there first appears an element at which the elastic stress reaches the yield. We solve an elastic solution on commercial finite element software ANSYS (Academic Teaching Introductory, 14.0). Then, we find a friction angle such that a stress state first appears on the yield envelope, i.e.,  $g(\sigma_1, \sigma_3, \phi) = 0$. This friction angle is always larger than the actual structural failure. 

\begin{figure}[p!]
\begin{center}
\includegraphics[width=3in]{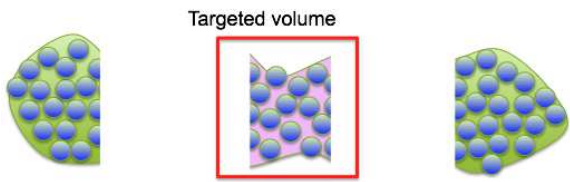}
\caption{Illustration of the upper bound computation associated with the partial volume stress. To investigate structural failure of the neck we consider the slice in the middle normal to the minor axis.}
\label{Fig:PartialVolume}
\end{center}
\end{figure}

On the other hand, the upper bound theorem states, ``if plastic deformation is assumed to be zero on the displacement boundary, then the loads determined by equating the rate at which the external forces do work to the rate of internal dissipation will be either higher than or equal to the actual limit load." We utilize the theorem by \cite{Holsapple2008A} that  guarantees the equivalence of the upper bound theorem and the yield condition of volume-average stresses. This paper uses two different types of volume-average stresses: the total volume stress and the partial volume stress. The total volume stress is the stress averaged over the whole body, which provides global failure of the body:
\begin{eqnarray}
\bar T^t_{ij} = \frac{1}{V} \int_V T_{ij} dv = \frac{1}{V} \int_V \rho x_j b_i dv. \label{Eq:TotalVolumeAverage}
\end{eqnarray}
where $i, j = 1,2,3$ and $(x_1, x_2, x_3) = (x,y,z)$. $V$ describes the whole volume. This stress was also discussed by \cite{Holsapple2008A}. For a real shape, however, the upper condition of this average stress is usually far away from the actual condition. For example, the stress state of a bifurcated body may be more complex than that of a spherical body. The partial volume stress, which is newly defined in this paper, is the stress averaged over an arbitrary slice normal to the minor principal axis (see Fig.\ref{Fig:PartialVolume}): 
\begin{eqnarray}
\bar T^p_{ij} &=&\frac{1}{V_p} \int_{V_p} \rho x_j b_i dv \nonumber \\ && + \frac{1}{V_p} \int_{S_p} l x_j T_{1i}  dx_2 dx_3, \label{Eq:PartialVolumeAverage}
\end{eqnarray} 
where $l$ is the direction cosine of the external normal to the cross section perpendicular to the $x$ axis. $l=1$ for the cross section opening to the positive direction, while $l=-1$ for that opening to the negative direction. The first term is an integral over the slice volume $V_p$, while the second term is an integral over the cross sections $S_p$. The upper bound condition by this stress is closer to the actual structural failure if a slice is properly chosen.

Computation of Eq.(\ref{Eq:TotalVolumeAverage}) and the first term of Eq.(\ref{Eq:PartialVolumeAverage}) is straightforward; however, that of the second term of Eq.(\ref{Eq:PartialVolumeAverage}) needs to be explained. If $j=1$, then $x_j$ can be treated as a constant value, and the integration becomes force balance on the cross sections. This procedure allows us to fix three stress components out of six stress components. Computational difficulty of this term appears when $j \ne 1$. This results from the the stress distribution on the cross sections. To avoid this difficulty, we assume that those three components are zero, using the fact that (216) Kleopatra is almost symmetric about the principal axes. After giving all stress components, we calculate the eigenvalues of this stress tensor to obtain the principal stresses. 

The partial volume stress will be used after the most sensitive slice to structural failure is determined. We search for the slice, considering a peak of the minimal principal axis component of normal stresses averaged over a cross section. The component is denoted as $\bar T^a_{11}$. \cite{Davidsson2001} proposed this stress average component to determine the failure condition of a biaxial body. Note that the area stress technique will be used only for finding the location of the most sensitive cross section, but not for determining structural failure. \cite{Sharma2009} reported that $\bar T^a_{11}$ is identical to the yield condition of the averaged normal stresses with $\phi=90^\circ$. 

\section{Results}
In this section, we show the results given in the range of the size scale from 1.0 to 1.5. In the following section, we call this range the test scale range. 

\subsection{Surface shedding}
\subsubsection{First shedding condition}
\begin{figure}[ht!]
	\begin{center}
		\subfigure[]{
         		\label{Fig:ZV5385}	
		\includegraphics[width=3in]{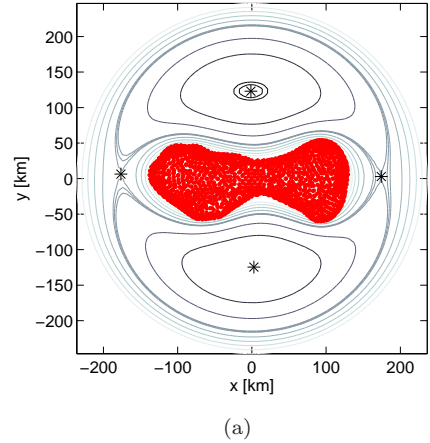}
          	}
		\subfigure[]{
         		\label{Fig:ZV281}	
		\includegraphics[width=3in]{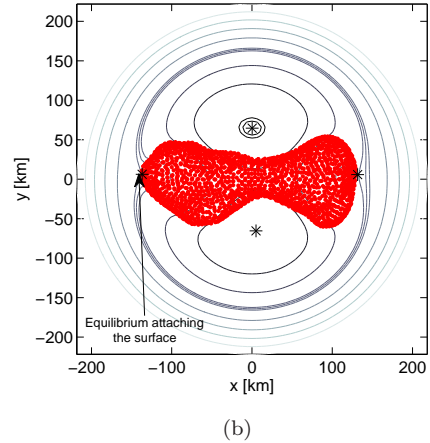}
          	} 	
	\caption{Zero-velocity curves for a size scale of 1.22, i.e., the \cite{Descamps2011} size. Figure \ref{Fig:ZV5385} shows the curves for the current spin period, i.e., 5.385 hr, and Fig.\ref{Fig:ZV281} describes those for a spin period of 2.81 hr at which the equilibrium point on the left reaches the surface.}
	\label{Fig:zeroVelocity}
	\end{center}
\end{figure}

As an example, we show the first shedding condition of (216) Kleopatra with a size of 1.22. Figure \ref{Fig:ZV5385} shows the zero-velocity curves at the current spin period, while Fig.\ref{Fig:ZV281} is for  the first shedding (2.85 hr). The red dots describe the shape projection onto the equatorial plane. The contour curves indicate the same energy levels, while the stars are the equilibrium points: two saddle points along the minor axis and two center points along the intermediate axis. In Fig.\ref{Fig:ZV5385}, a massless particle on the surface is in the primary's gravity dominant region and cannot lift off because none of the equilibrium points touches the surface. However, as this asteroid spins faster, the equilibrium points move to the surface, and the gravity dominant region shrinks. Eventually, as seen in Fig.\ref{Fig:ZV281}, the saddle point on the left side reaches the surface at 2.85 hr. In addition, the saddle point on the right side is also about to touch the surface. 

\begin{figure}[p!]
\begin{center}
\includegraphics[width=3in]{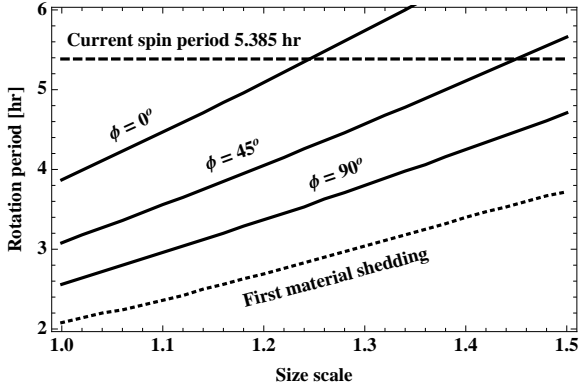}
\caption{First shedding (the dotted line) and the yield condition by the total volume stresses (the solid lines) as a function of the size scale. $\phi$ is the angle of friction. The first shedding is always below structural failure. }
\label{Fig:sheddingKleo}
\end{center}
\end{figure}

Figure \ref{Fig:sheddingKleo} indicates the relation between the first shedding condition (the dotted line) and the current spin period (the dashed line). To compare the first shedding with structural failure, we also plot the upper bound of structural failure of the whole volume (discussed later) at friction angles of $0^\circ$, $45^\circ$, and $90^\circ$ by solid lines. If the spin state is below these lines, the body must experience structural failure. It is found that the first shedding condition occurs with a much faster spin period than structural failure. Also, the current spin period is not faster than the first shedding condition in the test scale range. 

It is worth noting that although the first shedding is obtained numerically, the force balance between the gravity and the centrifugal force, i.e., $r_{Ostro} \alpha \Omega_{cr}^2 \propto GM/(r_{Ostro} \alpha)^2$, gives an analytical trend as 
\begin{eqnarray}
T_{cr} \propto \alpha^{3/2},
\end{eqnarray}
where $r_{Ostro}$ is the distance between the origin of the primary and the surface of the \cite{Ostro2000} size, $M$ is the mass of the primary, $\alpha$ is the size scale, $\Omega_{cr}$ is the first shedding, and $T_{cr} = 2\pi/\Omega_{cr}$.

\subsubsection{Hypothesis of satellites' origin}
As stated by \cite{Descamps2011}, the satellites are considered byproducts of a spin-up process leading to material shedding. Here, we use the technique for determining the first shedding; however, since we have already seen the zero-velocity curves of this asteroid earlier, we only track the location of the equilibrium points. The physical properties of these satellites are given in Table \ref{Table:satellite}. Also, it is assumed that their orbital planes are parallel to the equatorial plane of (216) Kleopatra, as given by \cite{Descamps2011}. 

The satellites are supposed to be small uniformed spheres. We call these satellites the test bodies. Initially, the test bodies are supposed to be located at the edges along the minor principal axis. The initial spin period is given by the conservation of the total angular momentum. Here, we neglect any other mass ejections' processes in the past for this consideration. On the assumption of zero eccentricity, the initial spin rate of the primary $\omega_0$ can be written as
\begin{eqnarray}
\omega_0 = \frac{I_z \omega_c + m_1 R_1 \Omega^2_1 + m_2 R_2 \Omega^2_2}{I_z + (m_1+m_2) r_{Ostro}^2 \alpha^2}, \label{Eq:newPeriod}
\end{eqnarray}
where $m_i$ and $R_i$ $(i=1,2)$ are the mass and the current distance from the center of mass of the primary, respectively. $I_z$ is the moment of inertia of the $z$ axis of the primary. $\omega_c$ is the current spin rate. Substitutions of the physical values on Table \ref{Table:satellite} into this equation determines the initial spin period as 5.086 hr. 

\begin{figure}[p!]
\begin{center}
\includegraphics[width=3in]{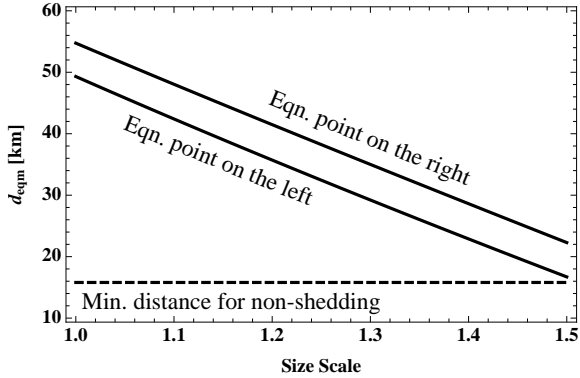}
\caption{Relation between the distance of the equilibrium points from the surface (the solid lines) and the minimal distance where material shedding does not occur, i.e., 15.8 km (the dashed line). This plot shows the case of a spin period of 5.086 hr. The period is obtained by Eq.(\ref{Eq:newPeriod}). The necessary condition of material shedding originating the satellites is that the equilibrium point on the left goes below the minimal distance.}
\label{Fig:scaleShedding}
\end{center}
\end{figure}

In this model, if the lifting condition satisfies, it is possible for the test bodies to initiate the satellites' formation. Technically, this condition is equivalent to that the saddle points are closer to the surface than the center of mass of the test bodies at this spin period. However, in this analysis, to make a stronger condition, we define that the test bodies lift off when the distance between the saddle points and the surface is less than the sum of these bodies' diameters, i.e., 15.8 km. Figure \ref{Fig:scaleShedding} shows the distance of the saddle points from the surface $d_{eqm}$ (the solid lines) and a distance of 15.8 km (the dashed line). Since, as shown earlier, surface shedding may occur on the left and right sides at almost the same rotation period, we track the distances of both points. The saddle point on the left side is always closer to the surface. The result shows that in the test scale range, the distances of the saddle points on both sides from the surface are never shorter than 15.8 km, and the test bodies cannot lift off the surface. It implies that the satellites do not result from simple fission of the original system, but may involve other processes such as the reaccumulation of an impact-generated debris disk.

\subsection{Structural failure as a function of size}
First, we discuss the lower bound. Since elastic solutions are independent of Young's modulus, we set the modulus as 10 GPa, which may be larger than usual geological materials on the Earth. On the other hand, different Poisson's ratios provide different solutions; therefore, we investigate the two cases: Poisson's ratio = 0.2 and 0.333. In the experiments, we investigate elastic solutions of 25 different size scales in the test scale range, i.e., $\alpha=1.00,1.02,1.04,...,1.50$. The result shows that in all the size scales, the solution includes elastic states which violates the MC condition even when $\phi=90^\circ$. It implies that (216) Kleopatra has plastic deformation of some small elements somewhere in all the test scale range. Again, this does not mean that it experiences plastic failure. 

\begin{figure}[ht!]
	\begin{center}
		\subfigure[]{
         		\label{Fig:elsFls100A}	
		\includegraphics[width=3in]{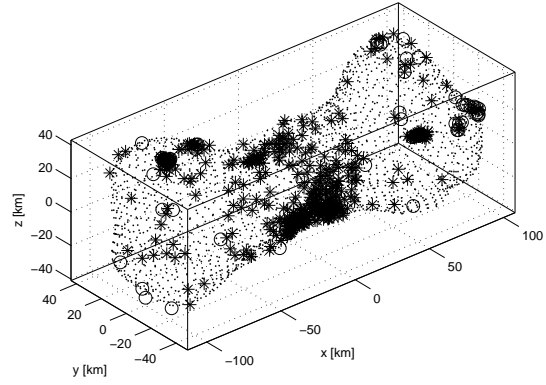}
          	}
		\subfigure[]{
         		\label{Fig:elsFls100B}	
		\includegraphics[width=3in]{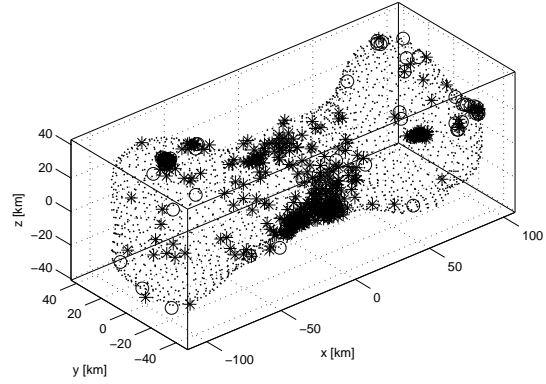}
          	} 	
	\caption{Elastic solutions for $\alpha=1.00$. The stars describes the stress states of which friction angle exceeds $50^\circ$. The circles mean that the stress states cannot be in the elastic region, even when the friction angle is $90^\circ$. The dots describe the shape of (216) Kleopatra. Figure \ref{Fig:elsFls100A} indicates the solution for Poisson's ratio = 0.2, while Fig.\ref{Fig:elsFls100B} shows the solution for Poisson's ratio = 0.333. It is found that different Poisson's ratios give different results, but they have the similar features. The stars mainly appear around the surface of the neck, while the circles are scattered on the whole surface.}
	\label{Fig:elsFls100}
	\end{center}
\end{figure}

\begin{figure}[ht!]
	\begin{center}
		\subfigure[]{
         		\label{Fig:elsFls130A}	
		\includegraphics[width=3in]{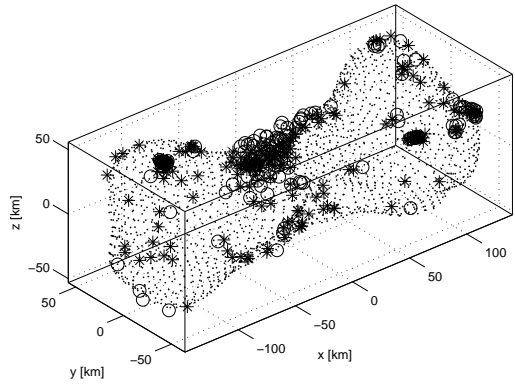}
          	}
		\subfigure[]{
         		\label{Fig:elsFls130B}	
		\includegraphics[width=3in]{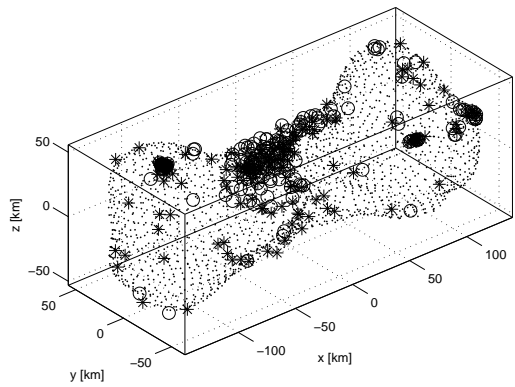}
          	} 	
	\caption{Elastic solutions for $\alpha=1.30$. We use the definitions given in Fig.\ref{Fig:elsFls100}. The stars assemble on the surface of the neck; however, in contrast to $\alpha=1.00$, their locations are the opposite side of the neck. The circles also appear near the stars. }
	\label{Fig:elsFls130}
	\end{center}
\end{figure}

\begin{figure}[ht!]
	\begin{center}
		\subfigure[]{
         		\label{Fig:elsFls150A}	
		\includegraphics[width=3in]{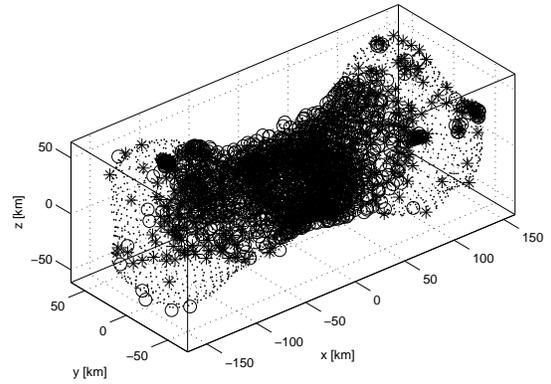}
          	}
		\subfigure[]{
         		\label{Fig:elsFls150B}	
		\includegraphics[width=3in]{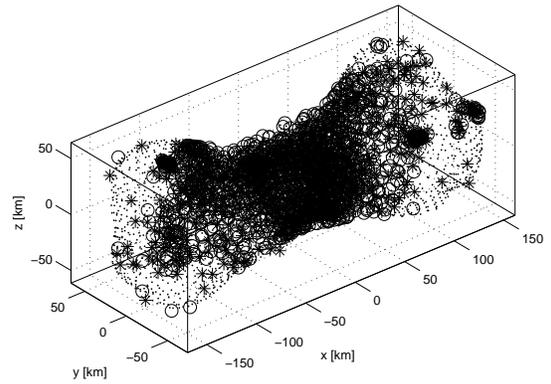}
          	} 	
	\caption{Elastic solutions for $\alpha=1.50$. Again, we use the definitions given in Fig.\ref{Fig:elsFls100}. In this case, the stars and circles spread out the whole neck. }
	\label{Fig:elsFls150}
	\end{center}
\end{figure}

Figures \ref{Fig:elsFls100} through \ref{Fig:elsFls150} show the elastic solutions which exceed $\phi=50^\circ$ (the stars) and those which cannot be in the elastic region even when $\phi=90^\circ$ (the circles). The dots indicate the shape of (216) Kleopatra. In addition, these figures describe the cases $\alpha=1.00$, $\alpha=1.30$, and $\alpha=1.50$, respectively. Each case is shown by the two Poisson's ratios: (a) 0.2 and (b) 0.333. It is found that in all the cases, although different Poisson's ratios give different solutions, the results are not significantly different. In Fig.\ref{Fig:elsFls100}, there are the stars around the surface of the neck, while the circles are scattered on the whole surface. In Fig.\ref{Fig:elsFls130}, the stars appear around the surface of the neck, but on the opposite side of $\alpha=1.00$. On the other hand, as shown in Fig.\ref{Fig:elsFls150}, the case $\alpha=1.50$ indicates that the stars and the circles are condensed around the neck. From this analysis, it is found that (216) Kleopatra is always above the lower bound of structural failure in the test scale range.

\begin{figure}[p!]
\begin{center}
\includegraphics[width=3in]{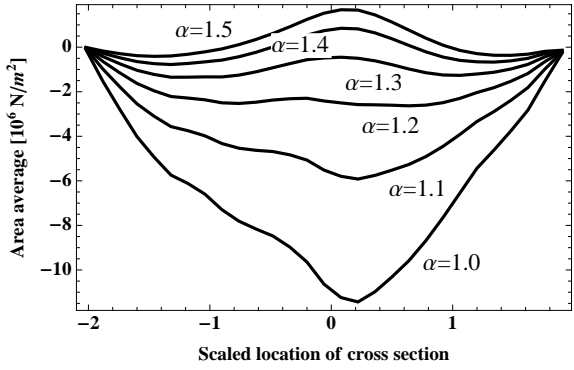}
\caption{The minor axis component of the averaged normal stresses as a function of locations of the normal cross sections. The negative value is compression, while the positive value is tension. Note that using the normalized length, we adjust the locations of the cross section so that the bodies with different size scales match their edges equally. The normalized length is given dividing the location by the equivalent radius. When $\alpha=1.3$, the part around the neck starts experiencing zero-tension, which is the most sensitive to structural failure.}
\label{Fig:plotScaleKleo}
\end{center}
\end{figure}

The upper bound condition of the partial volume is calculated by the following. The most sensitive cross section to structural failure is evaluated by using $\bar T^a_{11}$. Figure \ref{Fig:plotScaleKleo} shows the stress component in the vertical axis and a scaled length normalized by the equivalent radius in the horizontal axis. The body size along the minor principal axis ranges from $ - 2.1$ to $1.9$. Later, any lengths are introduced by the normalized length. When $\alpha=1.00$, since the body density is large, the body force is dominated by the self-gravity. As the size scale increases, however, the magnitude of the centrifugal force increases, and there appears a stress peak in the middle. At $\alpha=1.30$, the peak reaches the zero-tension, and the tension region starts spreading over the body. This result implies that the neck part located in the middle is quite sensitive to structural failure. In the following, to focus on the neck, we define the partial volume by the cuts at -0.21 and 0.58. 

\begin{figure}[p!]
\begin{center}
\includegraphics[width=3in]{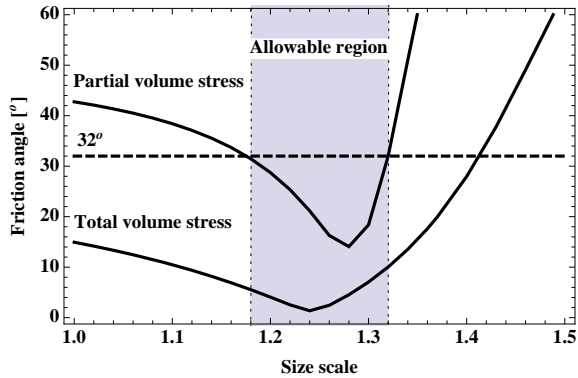}
\caption{Minimal friction angles of the total volume stress and partial volume stress (the solid lines) and the assumed Kleopatra's friction angle $32^\circ$ (the dashed line). The shaded region where the actual friction angle is larger than the minimal friction angle associated with the partial volume stress is allowable for the existence of the current shape.}
\label{Fig:MCwithStress}
\end{center}
\end{figure}

Since (216) Kleopatra is considered a cohesionless body, only the angle of friction is a free parameter. Here, we obtain the friction angle, using the upper bound techniques. This friction angle is identical to the minimal friction angle that the body can keep the original shape. More precisely, if the actual friction angle of the total (partial) volume is lower than the minimal friction angle, the total (partial) volume should fail. Since the most sensitive part is chosen as the partial volume, the actual friction angle should be always above the minimal friction angle of the partial volume. Figure \ref{Fig:MCwithStress} shows the minimal friction angles of the total volume and that of the partial volume as functions of the size scale $\alpha$. The minimal friction angle of the total volume keeps small angles in small scales and increases gradually. Also, when the size scale is 1.24, there is the minimal value $\sim 1^\circ$. On the other hand, the minimal friction angle of the partial volume is relatively high in a small scale, $\sim 43^\circ$ at $\alpha=1.0$, but small around the middle, $\sim 14^\circ$ at $\alpha=1.28$. Then, when the size scale is larger than 1.28, this friction angle increases dramatically. Since the minimal friction angle of the partial volume is always larger than that of the total volume, it can be concluded that the neck part is more sensitive to structural failure than the whole body.

\section{Discussion}
Before the discussion, we introduce a possible friction angle of (216) Kleopatra. \cite{Scott1963} showed that friction angles depend on materials' porosity (see Fig.7-5(a-c) on p.309). On the other hand, according to \cite{Ostro2000}, (216) Kleopatra's surface properties are comparable to lunar soil. They also stated that the estimated bulk density 3.5 g/cm$^3$ is consistent with either a solid enstatite-chondritic surface or a metallic surface with porosity of $< $ 60$\%$\footnote{As discussed earlier, given the \cite{Descamps2011} mass, the \cite{Ostro2000} density may be more than 6.0 g/cm$^3$.}. We assume this asteroid's porosity as $44\%$, the mean value of lunar soil's porosity ranging from $33\%$ and $55\%$.  Figure 7-5(a) on p.309 by \cite{Scott1963} shows that at a porosity of $44\%$, an allowable friction angle is $32^\circ$. We suppose that the body has a uniform structure as well. From these facts and assumptions, in the following discussion, the friction angle of (216) Kleopatra is fixed as 32$^\circ$. This implies that the body fails when the minimal friction angle is greater than 32$^\circ$.

Interestingly, Fig.\ref{Fig:MCwithStress} reveals that the allowable size scale lies between 1.18 and 1.32. This fact implies that if the size scale is not in this range, (216) Kleopatra cannot hold the current neck part. At a friction angle of 15$^\circ$, the lowest minimal friction angle of the partial volume, where the size scale is 1.15,  this asteroid encounters the most relaxed configuration. However, it does not mean that this asteroid' stress should be at this point. The important point here is that only the range between 1.18 and 1.32 is structurally allowable. Since the bulk density is from 2.9 to 3.8 g/cm$^3$, our estimation is consistent with \cite{Ostro2000}'s surface density estimation and with \cite{Descamps2011}'s bulk density estimation. From this result, the nominal size estimated by \cite{Ostro2000} is somewhat small, while that by \cite{Marchis2012} is relatively large. On the other hand, the estimation by \cite{Descamps2011} corresponds to our size evaluation. Note that their error estimations involve our result. If the size is assumed to be the \cite{Descamps2011} size, the minimal friction angle is no less than 27$^\circ$, which is within usual friction angles of geological materials (from 30$^\circ$ to 45$^\circ$). 

(216) Kleopatra may be sitting near plastic structural failure at the current spin period because the stable region for the current shape, especially the neck, is relatively small. \cite{Pravec2007} showed the spin barrier as a function of lightcurve amplitude (which is a proxy for asteroid equatorial elongation). Especially, in Fig.2 in their paper, the theoretical curves defined for ellipsoidal figures and for a friction angle of 90$^\circ$ indicate that the barrier shifts to lower spin rates for complex shapes and lower friction angles. Our result is consistent with their interpretation. Therefore, the neck may play an important role in sustaining the whole body; this asteroid may not be a contact binary composed of two bodies that loosely rest on each other. It means that this part might have been stretched plastically to get the current shape. It can be imagined that the narrow neck part is the byproduct of plastic deformation, and the original shape should have a wider neck and be less elongated than the current shape. We emphasize anew that a YORP effect and tidal perturbation do not have a significant effect on the spin state change of this asteroid. 

In contrast to plastic failure, material shedding can not occur at the current spin period. As shown in Fig.\ref{Fig:sheddingKleo}, material shedding does not happen even if this asteroid spins up to the condition where a material with a friction angle of $90^\circ$ encounters structural failure of the whole body. In addition, Fig.\ref{Fig:scaleShedding} shows that the distance between the surface and the equilibrium points does not reach the size of these satellites in the test scale range. 

\section{Conclusion}
This paper explored the dynamical and structural stability of the shape of Asteroid (216) Kleopatra at the current spin period, varying the shape size. We investigated the material shedding condition and the structural failure condition, separately. To find the condition where material shedding occurs first, we constructed the zero-velocity surfaces to find the dynamical equilibrium points and the constraints of the motion. The result shows that (216) Kleopatra cannot experience material shedding at the current spin period, and the satellites orbiting about the primary do not result from the shedding process. On other hand, to determine the lower and upper bounds of structural failure, we utilized limit analysis. It is found that the body, especially the neck part, is very sensitive to structural failure. Using elastic solutions, we revealed that (216) Kleopatra is always above the lower bound in the test scale range. Referring to \cite{Scott1963} to determine the friction angle of (216) Kleopatra as $32^\circ$, we found that only the size scale between 1.18 and 1.32 allows the body to be structurally stable. Our study agreed with the \cite{Descamps2011} size estimation. 



\acknowledgments

The authors thank Dr. Keith A. Holsapple for his dedicated technical advice. The authors also appreciate Dr. Pascal Descamps for the information about the estimation for the orbits of the satellites, Dr. Petr Pravec for useful discussion about the spin barrier for large elongated asteroids, Dr. Mikko Kaasalainen for useful discussion about their model of Kleopatra, and Dr. Franck Marchis for useful discussion about their estimation techniques. 





\appendix

\section{Comparison of the equilibrium points between \cite{Yu2012, Yu2013} and our computation} 

\label{App:Comp}
We show comparison between \cite{Yu2012,Yu2013} and our computation in Table \ref{Table:Comp}. Note that we recalculate their results of the equilibrium points by using our code, so these values are slightly different from those by \cite{Yu2012,Yu2013}. This difference results from computational thresholds in our code. As mentioned in the main text, we assume that the mass is fixed as $4.64 \times 10^{18}$ kg. If we choose the \cite{Ostro2000} size, the density should be $\sim$ 6.5 g/cm$^3$. Therefore, the equilibrium points by our code are farther away from the surface than those by \cite{Yu2012,Yu2013}.

\bibliographystyle{model2-names}
\bibliography{refInterior}  

\begin{thebibliography}{45}
\expandafter\ifx\csname natexlab\endcsname\relax\def\natexlab#1{#1}\fi
\expandafter\ifx\csname url\endcsname\relax
  \def\url#1{\texttt{#1}}\fi
\expandafter\ifx\csname urlprefix\endcsname\relax\def\urlprefix{URL }\fi
\providecommand{\eprint}[2][]{\url{#2}}
\providecommand{\bibinfo}[2]{#2}
\ifx\xfnm\relax \def\xfnm[#1]{\unskip,\space#1}\fi
\bibitem[{Bus and Binzel(2002)}]{Bus2002}
\bibinfo{author}{Bus, S.J.}, \bibinfo{author}{Binzel, R.P.},
  \bibinfo{year}{2002}.
\newblock \bibinfo{title}{Phase ii of the small main-belt asteroid
  spectroscopic survey: A feature-based taxonomy}.
\newblock \bibinfo{journal}{Icarus} \bibinfo{volume}{158},
  \bibinfo{pages}{146--177}.
\bibitem[{Cellino et~al.(1985)Cellino, Pannunzio, Zappala, Farinella and
  Paolicchi}]{Cellino1985}
\bibinfo{author}{Cellino, A.}, \bibinfo{author}{Pannunzio, R.},
  \bibinfo{author}{Zappala, V.}, \bibinfo{author}{Farinella, P.},
  \bibinfo{author}{Paolicchi, P.}, \bibinfo{year}{1985}.
\newblock \bibinfo{title}{Do we observe light curves of binary asteroids?}
\newblock \bibinfo{journal}{Astronomy and Astrophysics} \bibinfo{volume}{144},
  \bibinfo{pages}{355--362}.
\bibitem[{Chakrabarty(2006)}]{Chakrabarty2006}
\bibinfo{author}{Chakrabarty, J.}, \bibinfo{year}{2006}.
\newblock \bibinfo{title}{Theory of Plasticity}.
\newblock \bibinfo{publisher}{Elsevier Butterworth-Heinemann}.
  \bibinfo{edition}{third edition} edition.
\bibitem[{Chen and Han(1988)}]{Chen1988}
\bibinfo{author}{Chen, W.F.}, \bibinfo{author}{Han, D.J.},
  \bibinfo{year}{1988}.
\newblock \bibinfo{title}{Plasticity for Structural Engineers}.
\newblock \bibinfo{publisher}{Springer-Verlag}.
\bibitem[{Chree(1889)}]{Chree1889}
\bibinfo{author}{Chree, C.}, \bibinfo{year}{1889}.
\newblock \bibinfo{title}{The equations of an isotropic elastic solid in polar
  and cylindrical co-ordinates their solution and application}.
\newblock \bibinfo{journal}{Transactions of the Cambridge Philosophical
  Society} \bibinfo{volume}{14}, \bibinfo{pages}{250}.
\bibitem[{Chree(1891)}]{Chree1891}
\bibinfo{author}{Chree, C.}, \bibinfo{year}{1891}.
\newblock \bibinfo{title}{Xxxiii. some applications of physics and mathematics
  to geology}.
\newblock \bibinfo{journal}{The London, Edinburgh, and Dublin philosophical
  magazine and journal of science} \bibinfo{volume}{32}.
\bibitem[{Davidsson(2001)}]{Davidsson2001}
\bibinfo{author}{Davidsson, B.J.R.}, \bibinfo{year}{2001}.
\newblock \bibinfo{title}{Tidal splitting and rotational breakup of solid
  biaxial ellipsoids}.
\newblock \bibinfo{journal}{Icarus} \bibinfo{volume}{149},
  \bibinfo{pages}{375--383}.
\bibitem[{Descamps et~al.(2011)Descamps, Marchis, Berthier, Emery, Duch{\^e}ne,
  de~Pater, Wongb, Lim, Hammel, Vachier, Wiggins, Teng-Chuen-Yu, Peyrot,
  Pollock, Assafin, Vieira-Martins, Camargo, Braga-Ribas and
  Macomber}]{Descamps2011}
\bibinfo{author}{Descamps, P.}, \bibinfo{author}{Marchis, F.},
  \bibinfo{author}{Berthier, J.}, \bibinfo{author}{Emery, J.},
  \bibinfo{author}{Duch{\^e}ne, G.}, \bibinfo{author}{de~Pater, I.},
  \bibinfo{author}{Wongb, M.}, \bibinfo{author}{Lim, L.},
  \bibinfo{author}{Hammel, H.}, \bibinfo{author}{Vachier, F.},
  \bibinfo{author}{Wiggins, P.}, \bibinfo{author}{Teng-Chuen-Yu, J.P.},
  \bibinfo{author}{Peyrot, A.}, \bibinfo{author}{Pollock, J.},
  \bibinfo{author}{Assafin, M.}, \bibinfo{author}{Vieira-Martins, R.},
  \bibinfo{author}{Camargo, J.}, \bibinfo{author}{Braga-Ribas, F.},
  \bibinfo{author}{Macomber, B.}, \bibinfo{year}{2011}.
\newblock \bibinfo{title}{Triplicity and physical characteristics of asteroid
  (216) kleopatra}.
\newblock \bibinfo{journal}{Icarus} \bibinfo{volume}{211},
  \bibinfo{pages}{1022--1033}.
\bibitem[{Dobrovolskis(1982)}]{Dobrovolskis1982}
\bibinfo{author}{Dobrovolskis, A.R.}, \bibinfo{year}{1982}.
\newblock \bibinfo{title}{Internal stresses in phobs and other triaxial
  bodies}.
\newblock \bibinfo{journal}{Icarus} \bibinfo{volume}{52},
  \bibinfo{pages}{136--148}.
\bibitem[{Dunham et~al.(1991)Dunham, Osborn, Williams, Brisbin, Gada, Hirose,
  Maley, Povenmire, Stamm, Thrush, Aikman, Fletcher, Soma and
  Sichao}]{Dunham1991}
\bibinfo{author}{Dunham, D.W.}, \bibinfo{author}{Osborn, W.},
  \bibinfo{author}{Williams, G.}, \bibinfo{author}{Brisbin, J.},
  \bibinfo{author}{Gada, A.}, \bibinfo{author}{Hirose, T.},
  \bibinfo{author}{Maley, P.}, \bibinfo{author}{Povenmire, H.},
  \bibinfo{author}{Stamm, J.}, \bibinfo{author}{Thrush, J.},
  \bibinfo{author}{Aikman, C.}, \bibinfo{author}{Fletcher, M.},
  \bibinfo{author}{Soma, M.}, \bibinfo{author}{Sichao, W.},
  \bibinfo{year}{1991}.
\newblock \bibinfo{title}{The sizes and shapes of (4) vesta, (216) kleopatra,
  and (381) myrrha from occultations observed during january 1991}, in:
  \bibinfo{booktitle}{asteroids comets, meteors}, p.~\bibinfo{pages}{54}.
\bibitem[{Eriguchi et~al.(1982)Eriguchi, Hachisu and Sugimoto}]{Eriguchi1982}
\bibinfo{author}{Eriguchi, Y.}, \bibinfo{author}{Hachisu, I.},
  \bibinfo{author}{Sugimoto, D.}, \bibinfo{year}{1982}.
\newblock \bibinfo{title}{Dumb-bell-shape equilibria and mass-shedding
  pear-shape of selfgravitating incompressible fuild}.
\newblock \bibinfo{journal}{Progress of Theoretical Physics}
  \bibinfo{volume}{67}, \bibinfo{pages}{1068--1075}.
\bibitem[{Guibout and Scheeres(2003)}]{Guibout2003}
\bibinfo{author}{Guibout, V.}, \bibinfo{author}{Scheeres, D.J.},
  \bibinfo{year}{2003}.
\newblock \bibinfo{title}{Stability of surface motion on a rotating ellipsoid}.
\newblock \bibinfo{journal}{Celestial Mechanics and Dynamical Asronomy}
  \bibinfo{volume}{87}, \bibinfo{pages}{263--290}.
\bibitem[{Hestroffer et~al.(2002a)Hestroffer, Berthier, Descamps, Tanga,
  Cellino, Lattanzi, Martino and Zappala}]{Hestroffer2002A}
\bibinfo{author}{Hestroffer, D.}, \bibinfo{author}{Berthier, J.},
  \bibinfo{author}{Descamps, P.}, \bibinfo{author}{Tanga, P.},
  \bibinfo{author}{Cellino, A.}, \bibinfo{author}{Lattanzi, M.},
  \bibinfo{author}{Martino, M.D.}, \bibinfo{author}{Zappala, V.},
  \bibinfo{year}{2002}a.
\newblock \bibinfo{title}{Asteroid (216) kleopatra tests of the radar-derived
  shape model}.
\newblock \bibinfo{journal}{Astronomy and Astrophysics} \bibinfo{volume}{392},
  \bibinfo{pages}{729--733}.
\bibitem[{Hestroffer et~al.(2002b)Hestroffer, Marchis, Fusco and
  Berthier}]{Hestroffer2002B}
\bibinfo{author}{Hestroffer, D.}, \bibinfo{author}{Marchis, F.},
  \bibinfo{author}{Fusco, T.}, \bibinfo{author}{Berthier, J.},
  \bibinfo{year}{2002}b.
\newblock \bibinfo{title}{Adaptive optics observations of asteroid (216)
  kleopatra}.
\newblock \bibinfo{journal}{Astronomy and Astrophysics} \bibinfo{volume}{394},
  \bibinfo{pages}{339--343}.
\bibitem[{Holsapple(2004)}]{Holsapple2004}
\bibinfo{author}{Holsapple, K.A.}, \bibinfo{year}{2004}.
\newblock \bibinfo{title}{Equilibrium figures of spinning bodies with
  self-gravity}.
\newblock \bibinfo{journal}{Icarus} \bibinfo{volume}{172},
  \bibinfo{pages}{272--303}.
\bibitem[{Holsapple(2007)}]{Holsapple2007}
\bibinfo{author}{Holsapple, K.A.}, \bibinfo{year}{2007}.
\newblock \bibinfo{title}{Spin limits of solar system bodies: From the small
  fast-rotators to 2003 el61}.
\newblock \bibinfo{journal}{Icarus} \bibinfo{volume}{187},
  \bibinfo{pages}{500--509}.
\bibitem[{Holsapple(2008)}]{Holsapple2008A}
\bibinfo{author}{Holsapple, K.A.}, \bibinfo{year}{2008}.
\newblock \bibinfo{title}{Spinning rods, elliptical disks and solid ellipsoidal
  bodies: Elastic and plastic stresses and limit spins}.
\newblock \bibinfo{journal}{International journal of Non-Linear Mechanics}
  \bibinfo{volume}{43}, \bibinfo{pages}{733--742}.
\bibitem[{Holsapple(2010)}]{Holsapple2010}
\bibinfo{author}{Holsapple, K.A.}, \bibinfo{year}{2010}.
\newblock \bibinfo{title}{On yorp-induced spin deformations of asteroids}.
\newblock \bibinfo{journal}{Icarus} \bibinfo{volume}{205},
  \bibinfo{pages}{430--442}.
\bibitem[{Hosapple(2001)}]{Holsapple2001}
\bibinfo{author}{Hosapple, K.A.}, \bibinfo{year}{2001}.
\newblock \bibinfo{title}{Equilibrium configurations of solid cohesionless
  bodies}.
\newblock \bibinfo{journal}{Icarus} \bibinfo{volume}{154},
  \bibinfo{pages}{432--448}.
\bibitem[{Kaasalainen and Viikinkoski(2012)}]{Kaasalainen2012}
\bibinfo{author}{Kaasalainen, M.}, \bibinfo{author}{Viikinkoski, M.},
  \bibinfo{year}{2012}.
\newblock \bibinfo{title}{Shape reconstruction of irregular bodies with
  complementary data sources}.
\newblock \bibinfo{journal}{Astronomy and Astrophysics} \bibinfo{volume}{543},
  \bibinfo{pages}{A97}.
\bibitem[{Kadish et~al.(2008)Kadish, Barber, Washabaugh and
  Scheeres}]{Kadish2008}
\bibinfo{author}{Kadish, J.}, \bibinfo{author}{Barber, J.},
  \bibinfo{author}{Washabaugh, P.D.}, \bibinfo{author}{Scheeres, D.J.},
  \bibinfo{year}{2008}.
\newblock \bibinfo{title}{Stresses in accreted planetary bodies}.
\newblock \bibinfo{journal}{Solids and structures} \bibinfo{volume}{45},
  \bibinfo{pages}{540--550}.
\bibitem[{Love(1944)}]{Love1944}
\bibinfo{author}{Love, A.E.H.}, \bibinfo{year}{1944}.
\newblock \bibinfo{title}{A Treatise on the Mathematical Theory of Elasticity}.
\newblock \bibinfo{publisher}{Cambridge University Press}.
\bibitem[{Magnusson(1990)}]{Magnusson1990}
\bibinfo{author}{Magnusson, P.}, \bibinfo{year}{1990}.
\newblock \bibinfo{title}{Spin vectors of 22 large asteroids}.
\newblock \bibinfo{journal}{Icarus} \bibinfo{volume}{85},
  \bibinfo{pages}{229--240}.
\bibitem[{Marchis et~al.(2012)Marchis, Enriquez, Emery, Mueller, Baek, Pollock,
  Assafin, Martins, Berthier, Vachier, Cruikshank, Lim, Reichart, Ivarsen,
  Haislip and LaCluyze}]{Marchis2012}
\bibinfo{author}{Marchis, F.}, \bibinfo{author}{Enriquez, J.},
  \bibinfo{author}{Emery, J.}, \bibinfo{author}{Mueller, M.},
  \bibinfo{author}{Baek, M.}, \bibinfo{author}{Pollock, J.},
  \bibinfo{author}{Assafin, M.}, \bibinfo{author}{Martins, R.V.},
  \bibinfo{author}{Berthier, J.}, \bibinfo{author}{Vachier, F.},
  \bibinfo{author}{Cruikshank, D.}, \bibinfo{author}{Lim, L.},
  \bibinfo{author}{Reichart, D.}, \bibinfo{author}{Ivarsen, K.},
  \bibinfo{author}{Haislip, J.}, \bibinfo{author}{LaCluyze, A.},
  \bibinfo{year}{2012}.
\newblock \bibinfo{title}{Multiple asteroid systems: Dimensions and thermal
  properties from spitzer space telescope and ground-based observations}.
\newblock \bibinfo{journal}{Icarus} \bibinfo{volume}{221},
  \bibinfo{pages}{1130--1161}.
\bibitem[{Mitchell et~al.(1995)Mitchell, Ostro and Rosema}]{Mitchell1995}
\bibinfo{author}{Mitchell, D.L.}, \bibinfo{author}{Ostro, S.J.},
  \bibinfo{author}{Rosema, K.D.}, \bibinfo{year}{1995}.
\newblock \bibinfo{title}{Radar observations of asteroid 7 iris, 9 metis, 12
  victoria, 216 kleopatra, and 654 zelinda}.
\newblock \bibinfo{journal}{Icarus} \bibinfo{volume}{118},
  \bibinfo{pages}{105--131}.
\bibitem[{Ostro et~al.(2000)Ostro, Hudson, Nolan, Margot, Scheeres, Campbell,
  Magri, Giorgini and Yeomans}]{Ostro2000}
\bibinfo{author}{Ostro, S.J.}, \bibinfo{author}{Hudson, R.S.},
  \bibinfo{author}{Nolan, M.C.}, \bibinfo{author}{Margot, J.L.},
  \bibinfo{author}{Scheeres, D.J.}, \bibinfo{author}{Campbell, D.B.},
  \bibinfo{author}{Magri, C.}, \bibinfo{author}{Giorgini, J.D.},
  \bibinfo{author}{Yeomans, D.K.}, \bibinfo{year}{2000}.
\newblock \bibinfo{title}{Radar observations of asteroid 216 kleopatra}.
\newblock \bibinfo{journal}{Science} \bibinfo{volume}{288},
  \bibinfo{pages}{836--839}.
\bibitem[{Ostro et~al.(2006)Ostro, Margot, Benner, Giorgini, Scheeres,
  Fahnestock, Broschart, Bellerose, Nolan, Magri, Pravec, Scheirich, Rose,
  Jurgens, Jong and Suzuki}]{Ostro2006}
\bibinfo{author}{Ostro, S.J.}, \bibinfo{author}{Margot, J.L.},
  \bibinfo{author}{Benner, L.A.M.}, \bibinfo{author}{Giorgini, J.D.},
  \bibinfo{author}{Scheeres, D.J.}, \bibinfo{author}{Fahnestock, E.G.},
  \bibinfo{author}{Broschart, S.B.}, \bibinfo{author}{Bellerose, J.},
  \bibinfo{author}{Nolan, M.C.}, \bibinfo{author}{Magri, C.},
  \bibinfo{author}{Pravec, P.}, \bibinfo{author}{Scheirich, P.},
  \bibinfo{author}{Rose, R.}, \bibinfo{author}{Jurgens, R.F.},
  \bibinfo{author}{Jong, E.M.D.}, \bibinfo{author}{Suzuki, S.},
  \bibinfo{year}{2006}.
\newblock \bibinfo{title}{Radar imaging of binary near-earth asteroid (66391)
  1999kw4}.
\newblock \bibinfo{journal}{Science} \bibinfo{volume}{314},
  \bibinfo{pages}{1276--1280}.
\bibitem[{Pravec et~al.(2007)Pravec, Harris and Warner}]{Pravec2007}
\bibinfo{author}{Pravec, P.}, \bibinfo{author}{Harris, A.W.},
  \bibinfo{author}{Warner, B.D.}, \bibinfo{year}{2007}.
\newblock \bibinfo{title}{Nea rotations and binaries}.
\newblock \bibinfo{journal}{Near Earth Objects, our Celestial Neighbors:
  Opportunity and Risk, Proceedings IAU Symposium} .
\bibitem[{Sanchez and Scheeres(2012)}]{Sanchez2012}
\bibinfo{author}{Sanchez, P.}, \bibinfo{author}{Scheeres, D.J.},
  \bibinfo{year}{2012}.
\newblock \bibinfo{title}{Dem simulation of rotation-induced reshaping and
  disruption of rubble-pile asteroids}.
\newblock \bibinfo{journal}{Icarus} .
\bibitem[{Scaltriti and Zappala(1978)}]{Scaltriti1978}
\bibinfo{author}{Scaltriti, F.}, \bibinfo{author}{Zappala, V.},
  \bibinfo{year}{1978}.
\newblock \bibinfo{title}{Photoelectric photometry of asteroids 37, 80, 97,
  216, 270, 313, and 471}.
\newblock \bibinfo{journal}{Icarus} \bibinfo{volume}{34},
  \bibinfo{pages}{428--435}.
\bibitem[{Scott(1963)}]{Scott1963}
\bibinfo{author}{Scott, R.F.}, \bibinfo{year}{1963}.
\newblock \bibinfo{title}{Principles of Soil Mechanics}.
\newblock \bibinfo{publisher}{Addison-Wesley}.
\bibitem[{Sharma(2009)}]{Sharma2009}
\bibinfo{author}{Sharma, I.}, \bibinfo{year}{2009}.
\newblock \bibinfo{title}{The equilibrium of rubble-pile satellites: The darwin
  and roche ellipsoids for gravitationally held granular aggregates}.
\newblock \bibinfo{journal}{Icarus} \bibinfo{volume}{200},
  \bibinfo{pages}{636--654}.
\bibitem[{Sharma(2010)}]{Sharma2010}
\bibinfo{author}{Sharma, I.}, \bibinfo{year}{2010}.
\newblock \bibinfo{title}{Equilibrium shapes of rubble-pile binaries: The
  darwin ellipsoids for gravitationally held granular aggregates}.
\newblock \bibinfo{journal}{Icarus} \bibinfo{volume}{205},
  \bibinfo{pages}{638--657}.
\bibitem[{Takahashi et~al.(2004)Takahashi, Shinokawa, Yoshida, Mukai, Ip and
  Kawabata}]{Takahashi2004}
\bibinfo{author}{Takahashi, S.}, \bibinfo{author}{Shinokawa, K.},
  \bibinfo{author}{Yoshida, F.}, \bibinfo{author}{Mukai, T.},
  \bibinfo{author}{Ip, W.H.}, \bibinfo{author}{Kawabata, K.},
  \bibinfo{year}{2004}.
\newblock \bibinfo{title}{Photometric and polarimetric observations and model
  simulations of (216) kleopatra}.
\newblock \bibinfo{journal}{Earth Planets Space} \bibinfo{volume}{56},
  \bibinfo{pages}{997--1004}.
\bibitem[{Tanga et~al.(2001)Tanga, Hestroffer, Berthier, Cellino, Lattanzi,
  Martino and Zappala}]{Tanga2001}
\bibinfo{author}{Tanga, P.}, \bibinfo{author}{Hestroffer, D.},
  \bibinfo{author}{Berthier, J.}, \bibinfo{author}{Cellino, A.},
  \bibinfo{author}{Lattanzi, M.G.}, \bibinfo{author}{Martino, M.D.},
  \bibinfo{author}{Zappala, V.}, \bibinfo{year}{2001}.
\newblock \bibinfo{title}{Hst/fgs observations of the asteroid (216)
  kleopatra}.
\newblock \bibinfo{journal}{Icarus} \bibinfo{volume}{153},
  \bibinfo{pages}{451--454}.
\bibitem[{Tedesco et~al.(2002)Tedesco, Noah, Noah and and}]{Tedesco2002}
\bibinfo{author}{Tedesco, E.F.}, \bibinfo{author}{Noah, P.V.},
  \bibinfo{author}{Noah, M.}, \bibinfo{author}{and, S.D.P.},
  \bibinfo{year}{2002}.
\newblock \bibinfo{title}{The supplemental iras minor planet survey}.
\newblock \bibinfo{journal}{The astrophysical journal} \bibinfo{volume}{123},
  \bibinfo{pages}{1056--1085}.
\bibitem[{Tholen(1984)}]{Tholen1984}
\bibinfo{author}{Tholen, D.J.}, \bibinfo{year}{1984}.
\newblock \bibinfo{title}{Asteroid Taxonomy from Cluster Analysis of
  Photometry}.
\newblock Ph.D. thesis. The University of Arizona.
\bibitem[{Walsh et~al.(2008)Walsh, Richardson and Michel}]{Walsh2008}
\bibinfo{author}{Walsh, K.J.}, \bibinfo{author}{Richardson, D.C.},
  \bibinfo{author}{Michel, P.}, \bibinfo{year}{2008}.
\newblock \bibinfo{title}{Rotational breakup as the origin of small binary
  asteroids}.
\newblock \bibinfo{journal}{Nature} \bibinfo{volume}{454},
  \bibinfo{pages}{188--191}.
\bibitem[{Walsh et~al.(2012)Walsh, Richardson and Michel}]{Walsh2012}
\bibinfo{author}{Walsh, K.J.}, \bibinfo{author}{Richardson, D.C.},
  \bibinfo{author}{Michel, P.}, \bibinfo{year}{2012}.
\newblock \bibinfo{title}{Spin-up of rubble-pile asteroids: Disruption,
  satellite formation, and equilibrium shapes}.
\newblock \bibinfo{journal}{Icarus} \bibinfo{volume}{220},
  \bibinfo{pages}{514--529}.
\bibitem[{Washabaugh and Scheeres(2002)}]{Washabaugh2002}
\bibinfo{author}{Washabaugh, P.D.}, \bibinfo{author}{Scheeres, D.J.},
  \bibinfo{year}{2002}.
\newblock \bibinfo{title}{Energy and stress distributions in ellipsoids}.
\newblock \bibinfo{journal}{Icarus} \bibinfo{volume}{159},
  \bibinfo{pages}{314--321}.
\bibitem[{Weidenschilling(1980)}]{Weidenschilling1980}
\bibinfo{author}{Weidenschilling, S.J.}, \bibinfo{year}{1980}.
\newblock \bibinfo{title}{Hektor: Nature of origin of a binary asteroid}.
\newblock \bibinfo{journal}{Icarus} \bibinfo{volume}{44},
  \bibinfo{pages}{807--809}.
\bibitem[{Werner and Scheeres(1997)}]{Werner1996}
\bibinfo{author}{Werner, R.A.}, \bibinfo{author}{Scheeres, D.J.},
  \bibinfo{year}{1997}.
\newblock \bibinfo{title}{Exterior gravitation of polyhedron derived and
  compared with harmonic and mason gravitation represenations of asteroid 4769
  castalia}.
\newblock \bibinfo{journal}{Celestial Mechanics and Dynamical Asronomy}
  \bibinfo{volume}{65}, \bibinfo{pages}{313--344}.
\bibitem[{Yu and Baoyin(2012)}]{Yu2012}
\bibinfo{author}{Yu, Y.}, \bibinfo{author}{Baoyin, H.}, \bibinfo{year}{2012}.
\newblock \bibinfo{title}{Orbital dynamics in the vicinity of asteroid 216
  kleopatra}.
\newblock \bibinfo{journal}{The astronomical journal} \bibinfo{volume}{143},
  \bibinfo{pages}{62--71}.
\bibitem[{Yu and Baoyin(2013)}]{Yu2013}
\bibinfo{author}{Yu, Y.}, \bibinfo{author}{Baoyin, H.}, \bibinfo{year}{2013}.
\newblock \bibinfo{title}{Resonant orbits in the vicinity of asteroid 216
  kleopatra}.
\newblock \bibinfo{journal}{Astrophysics and Space Science}
  \bibinfo{volume}{343}, \bibinfo{pages}{74--82}.
\bibitem[{Zappala et~al.(1983)Zappala, Martino and Scaltriti}]{Zappala1983}
\bibinfo{author}{Zappala, V.}, \bibinfo{author}{Martino, M.D.},
  \bibinfo{author}{Scaltriti, F.}, \bibinfo{year}{1983}.
\newblock \bibinfo{title}{Photoeletric analysis of asteroid 216 kleopatra:
  Implications for its shape}.
\newblock \bibinfo{journal}{Icarus} \bibinfo{volume}{53},
  \bibinfo{pages}{458--464}.

\end{thebibliography}



\clearpage

\clearpage

\clearpage

\begin{table}
\begin{center}
\caption{Constant properties of (216) Kleopatra}
\label{Table:Kleopatra}
\begin{tabular}{c c c c}
\hline 
Property & Value & Units & Reference \\
\hline
Mass  & 4.64 $\times$ 10$^{18}$ & kg & \cite{Descamps2011} \\
Period & 5.385 & hr & \cite{Magnusson1990} \\
Cohesion & 0 & N$/$m$^3$ & \cite{Holsapple2007} \\
Shape & - & - & \cite{Ostro2000} \\
\hline
\end{tabular}
\end{center}
\end{table}

\begin{table}
\begin{center}
\caption{Physical properties of (216) Kleopatra's satellites. Note that the primary's spin pole is given by $\lambda=76\pm3^\circ$ and $\beta=16\pm1^\circ$ in J2000 ecliptic coordinates (\citealt{Descamps2011}). }
\label{Table:satellite}
\begin{tabular}{l l l}
\hline 
Property & Satellite (outer) & Satellite (inner) \\
\hline
Diameter [km] & $8.9\pm1.6$ & $6.9\pm1.6$ \\
Orbital period [days] & $2.32\pm0.02$ & $1.24\pm0.02$\\ 
Semi-major axis [km] & $678 \pm$13 & $454\pm6$ \\
Orbit pole right ascension [deg] & $74\pm2$ & $79\pm2$ \\
Orbit pole declination [deg] & $16\pm1$ & $16\pm1$ \\
\hline
\end{tabular}
\end{center}
\end{table}

\begin{table}
\begin{center}
\caption{Comparison of the equilibrium points by \cite{Yu2012, Yu2013} and our computations. Notations $E_i$ ($i=1,..,4)$ are based on Table 1 in \cite{Yu2012}. Note that we recalculate their results by using our code, but they are slightly different from their values in Table 1 in \cite{Yu2012}. This comes from convergent thresholds defined in our code.}
\label{Table:Comp}
\begin{tabular}{l l l l}
\hline 
\hline
Property & & \cite{Yu2012, Yu2013} & our computation \\
\hline
Volume [km$^3$] & & $7.09\times10^5$ & $7.09\times10^5$  \\
Density [g/cm$3$] & & $3.6$ & $6.5$\\ 
Mass [kg] & & $2.55\times10^{18}$ & $4.64\times10^{18}$ \\
\hline
\hline 
Equilibrium [km] & & & \\
\hline
& $x$ & $1.43\times10^2$ & $1.66\times10^2$ \\
$E_1$ & $y$ & $2.44$ & $2.27$ \\
& $z$ & $1.18$ & $7.91\times10^{-1}$ \\
& $x$ & $-1.45\times10^2$ & $-1.67\times10^2$ \\
$E_2$& $y$ & $5.19$ & $4.97$ \\
& $z$ & $-2.72\times10^{-1}$ & $-5.47\times10^{-2}$ \\
& $x$ & $2.22$ & $1.26$ \\
$E_3$& $y$ & $-1.02 \times 10^2$ & $-1.31 \times 10^2$ \\
& $z$ & $-2.72\times10^{-1}$ & $1.71 \times 10^{-1}$ \\
& $x$ & $-1.17$ & $-1.59$ \\
$E_4$& $y$ & $1.01\times10^2$ & $1.30 \times 10^2$ \\
& $z$ & $-5.46 \times 10^{-1}$ & $-3.32 \times 10^{-1}$ \\
\hline
\end{tabular}
\end{center}
\end{table}




\end{document}